\documentclass[12pt]{article}
 \usepackage{graphicx}  
\usepackage{amsmath}   
\usepackage[compress]{cite}
\usepackage{amssymb}   
\usepackage{bm} 
\usepackage{dcolumn}
\usepackage{ulem}
\usepackage{color}
\usepackage{mathtools}
\usepackage{mathrsfs}
\usepackage{amsfonts}
\usepackage{varioref}
\usepackage{textcomp}
\usepackage{geometry}
\usepackage{tcolorbox}
\usepackage{empheq}
\usepackage[active]{srcltx}
\usepackage{tikz}
\usepackage{lmodern}
\usepackage{braket}
\usetikzlibrary{decorations.pathmorphing}
\tikzset{snake it/.style={decorate, decoration=snake}}
\usepackage{fancybox}

\RequirePackage[colorlinks,citecolor=blue,urlcolor=magenta,linkcolor=blue]{hyperref}
\allowdisplaybreaks
 
 \def\nn{\nonumber}

\def\fft#1#2{{#1 \over #2}}

\def\be{\begin{equation}}
\def\bea{\begin{eqnarray}}
\def\ee{\end{equation}}
\def\eea{\end{eqnarray}}

\def\IR{\mathbb{R}}
\def\coeff#1#2{\relax{\textstyle {#1 \over #2}}\displaystyle}

\def\cF{{\cal F}}

\let\G=\Gamma
\let\s=\sigma \let\t=\tau
\let\d=\delta
\let\e=\varepsilon

\numberwithin{equation}{section}

 \title{\bf Killing Spinors for Finite Temperature \\ Euclidean Solutions  at the  BPS Bound}
 
\author{Subramanya Hegde$^1$ and  Amitabh Virmani$^2$}
\date{%
{\normalsize    $^1$The Institute of Mathematical Sciences \\ CIT Campus, Taramani,  Chennai 600 113, Tamil Nadu, India.\\%
\vskip 0.5cm    $^2$Chennai Mathematical Institute \\ H1, SIPCOT IT Park, Siruseri, Kelambakkam 603103, Tamil Nadu, India.} \\
\vskip 1cm    {\small subbuh@imsc.res.in, avirmani@cmi.ac.in} \\
\vskip 1cm    \today \\
}
\begin{document}
\maketitle 
\abstract{In a recent paper [arXiv:2308.00038], Anupam, Chowdhury, and Sen  conjectured that the finite temperature Euclidean five-dimensional Cvetic-Youm solution saturating the BPS bound is supersymmetric. In this paper,  we  explicitly construct Killing spinors for this solution in  five-dimensional minimal supergravity. We also expand on the previous discussions of Killing spinors for the finite temperature Euclidean Kerr-Newman solution saturating the BPS bound. For both these cases, we show that the total charge gets divided into two harmonic sources on three-dimensional flat base space. } 

\newpage

\tableofcontents

\section{Introduction}

Arguably, one of the biggest successes of string theory is the statistical mechanical explanation of the entropy of a class of supersymmetric black holes \cite{Strominger:1996sh, 0708.1270}. In order to compute this entropy in string theory, one typically considers a supersymmetric compactification. In such a set-up, at weak coupling the microstates are bound states of the fluctuations of strings and branes. For these states, one computes an appropriate supesymmetric helicity trace index. The index computes an integer which depends only on the charges of the black hole. The integral property of the index has been tested in great detail over the years \cite{Sen:2011ktd, Bringmann:2012zr, Chattopadhyaya:2017ews, 
Govindarajan:2022nzb}.

A key question in this subject has been whether this index can be computed using supergravity path integral? 
Earlier work used path integral in the near horizon geometry \cite{Sen:2008vm}. This provides a way to compare microscopic and macroscopic results. However, one encounters several difficulties in extending these computations beyond what has been computed so far \cite{Sen:2023dps}. A suggested way out of these difficulties is to do path integral  over the modes living in full asymptotically flat space-time geometry. The recent progress towards understanding the relationship between the supersymmetric helicity trace index and Euclidean gravity path integral  over the modes living in full asymptotically flat space-time geometry instead of just on the modes in the near horizon geometry
\cite{Cabo-Bizet:2018ehj, Heydeman:2020hhw, Iliesiu:2021are, Iliesiu:2022kny} has attracted a lot of attention and has led to intense activity in this area \cite{ Anupam:2023yns, Banerjee:2023quv, H:2023qko,
Chen:2023mbc, Kapec:2023ruw, Rakic:2023vhv, Boruch:2023gfn, Mondal:2023bcx}.

The classical limit of this relationship \cite{Iliesiu:2021are, Anupam:2023yns} leads to a non-trivial relation between the entropies of zero-temperature and finite temperature black holes. This has led to conjectures  that Euclidean finite temperature rotating black holes saturating the BPS bound are supersymmetric. These conjectures are somewhat surprising, as
it is generally believed that finite temperature metrics cannot admit Killing spinors, because the spinor field must be anti-periodic in imaginary time. However, the key point is, if the manifold is rotating, this naive intuition no longer applies. The contractible circle is a combination of the imaginary time and the angles. As a result, Killing spinors can and do exist. The main aim of this paper is to establish this for two examples: (i) the finite temperature Euclidean five-dimensional Cvetic-Youm solution saturating the BPS bound, (ii) the finite temperature Euclidean Kerr-Newman solution saturating the BPS bound. The case of Euclidean Kerr-Newman is best regarded as a review, as most aspects of this discussion are already well known in the literature. 
The case of Euclidean five-dimensional Cvetic-Youm solution was conjectured by Anupam, Chowdhury, and Sen (ACS) \cite{Anupam:2023yns}.\footnote{A related phenomenon was observed in \cite{Bobev:2020pjk} in the context of charged AdS black holes. Ref.~\cite{Bobev:2020pjk} highlights situations where there are no regular supersymmetric Lorentzian black holes with both electric and magnetic charges, but in the Euclidean setting the geometry is smooth and are valid Euclidean supersymmetric saddle points of the supergravity equations of motion. We thank Nikolay Bobev for discussions on this point.}

In such discussions,  Israel-Wilson-Perjes (IWP) metrics \cite{Perjes:1971gv, Israel:1972vx} and related constructions play key role.  IWP metrics admit Killing spinors \cite{Tod:1983pm}. Finite temperature Euclidean Kerr-Newman solution saturating the BPS bound can be written in the IWP form \cite{Israel:1972vx, Whitt:1984wk, Yuille:1987vw}.  Hence, it is clear that it admits Killing spinors.  In a recent paper,  Boruch et al \cite{Boruch:2023gfn} discuss IWP metrics for ${\cal N}=2$ supergravity coupled to arbitrary number of vector multiplets.

The case of five-dimensional Cvetic-Youm solution saturating the BPS bound is much less clear; certainly far from obvious. This case is our main focus. Along the way, we also discuss a class of supersymmetric solutions to Euclidean five-dimensional supergravity. 
One of our key technical observation is that the ACS metric can be conveniently written in the Chong, Cvetic, Lu, Pope (CCLP) form \cite{Chong:2005hr}. This observation makes the calculations manageable.

The rest of the paper is organised as follows. In section \ref{sec:EKN}, we start with a discussion 
Euclidean Kerr-Newman solution 
in pure ${\cal N}=2$ Euclidean supergravity. We also present some details on the supergravity theory and the electromagnetic duality highlighting various subtle minus signs that must be properly taken into account. In section \ref{sec:5d}, we discuss Killing spinors for the 
Euclidean 
Cvetic-Youm solution. 
In section \ref{sec:conclusions}, we end with our conclusions and a brief discussion of some of the open directions. Various technical details are relegated to three appendices.  Our gamma matrix and spinor conventions are spelled out in appendix \ref{app:gamma}.  
In appendix \ref{app:ACS}, we present a detailed comparison between the metric used in section \ref{sec:5d} and the metric given in the ACS paper.  For completeness, we also derive the pure ${\cal N}=2, D=4$  Euclidean supergravity in appendix \ref{app:conventions} using the superconformal formalism introduced in \cite{deWit:2017cle}.

\section{Euclidean Kerr-Newman solution}
\label{sec:EKN}

Although it appears to be well known that this solution is supersymmetric, it is difficult to find explicit Killing spinors in the literature. We fill this gap in this section. Our discussion here also sets the stage for five-dimensional case discussed in the next section. There is some overlap of our discussion 
with that of section 2 of the recent paper of Boruch et al \cite{Boruch:2023gfn}.

\subsection{Pure ${\cal N}=2$ Euclidean supergravity}
We start by recalling the essential aspects of pure ${\cal N}=2$ Euclidean supergravity \cite{Cortes:2009cs}. The bosonic sector of the theory consists of the vielbein $e_\mu^a$ and the graviphoton $A_\mu$. We are interested in the Killing spinor equation, and the bosonic equations of motion that follow as a result of integrability conditions of the Killing spinor equation. In appendix \ref{app:conventions}, we construct the Killing spinor equation as well as the supergravity Lagrangian from the superconformal approach developed in \cite{deWit:2017cle}. The integrability analysis is a review of the results in \cite{Dunajski:2010uv} with the cosmological constant set to zero.

The bosonic sector of pure Euclidean $\mathcal{N}=2$ supergravity is the Einstein-Maxwell theory with the equations of motion,
\begin{align}
R_{\mu \nu}+2c_1 \left(F_{\mu\rho}F_{\nu}{}^{\rho}-\frac{1}{4}F_{\rho \sigma} F^{\rho \sigma} g_{\mu\nu}\right)&=0,\nonumber\\
\nabla^\mu F_{\mu\nu}&=0,
\end{align}
where $c_1=\pm 1$ is an ambiguous sign  that reflects the ambiguity in the sign of the gauge kinetic term in four dimensional Euclidean Einstein-Maxwell theory \cite{Sabra:2016abd}. As shown in \cite{Sabra:2016abd},  theories with the two choices of sign are mapped to each other by electromagnetic duality. Let us briefly recall this. Consider the Lagrangian,
\begin{align}
S=\int d^4x \sqrt{g} \left(\frac{R}{4}-\frac{1}{4}F_{\mu \nu} F^{\mu \nu} + \frac{1}{2} b_\mu \partial_\nu F_{\rho\sigma}\epsilon^{\mu\nu\rho\sigma}\right), \label{lagrange_multiplier_b}
\end{align} 
where $b_\mu$ is a Lagrange multiplier whose equation of motion implements the Bianchi identity for the Maxwell field strength. To obtain the dual description, consider the equation of motion for the Maxwell field strength $F_{\rho\sigma}$ from  \eqref{lagrange_multiplier_b} which reads,
\begin{align}
F^{\mu\nu}=\frac{1}{2}\epsilon^{\mu\nu\rho\sigma}(\partial_\rho b_\sigma-\partial_\sigma b_\rho). \label{F-eq}
\end{align}
We then define $G_{\rho\sigma}=\partial_\rho b_\sigma-\partial_\sigma b_\rho$. Since \eqref{F-eq} is not a dynamical equation for the field strength $F_{\mu\nu}$, we can substitute for $F_{\mu\nu}$ in terms of $G_{\mu\nu}$ in \eqref{lagrange_multiplier_b} to obtain,
\begin{align}
S=\int d^4x \sqrt{g} \left(\frac{R}{4}+\frac{1}{16}\epsilon^{\mu\nu\rho\sigma}\epsilon_{\rho\sigma\alpha\beta}G_{\mu\nu}G^{\alpha\beta}\right) =\int d^4x \sqrt{g} \left(\frac{R}{4}+\frac{1}{4}G_{\mu\nu}G^{\mu\nu}\right),
\end{align}
where from the first line to the second we have used the property of Levi-Civita tensor in Euclidean space, $\epsilon^{\mu\nu\rho\sigma}\epsilon_{\rho\sigma\alpha\beta}=4\delta_{[\alpha}^\mu\delta_{\beta]}^\nu$. Thus, duality maps Maxwell Lagrangian with one signature for the kinetic term to the other. This leads to the ambiguity in the sign factor $c_1$ in the Einstein-Maxwell equations above. Given a solution to Einstein-Maxwell theory, it is important to check this sign, and consider the corresponding Killing spinor equation. In particular, for electric and magnetic solutions obtained by analytically continuation  of Lorentzian solutions via $t \to \pm \mathrm{i} \tau$, we need to consider Einstein-Maxwell equations with opposite $c_1$ signs.

To obtain the Killing spinor equation for the appropriate set-up, we need to consider the integrability condition for the Killing spinor and impose consistency with Einstein-Maxwell equations. This analysis is the same as \cite{Dunajski:2010uv}, and the Killing spinor equations read\footnote{For the spin-connection, we have taken the convention of \cite{Freedman:2012zz} (which are opposite to that of \cite{deWit:2017cle}). For us, $\omega(e)_\mu{\!}^{ab}=2\,e^{\nu[a}\partial_{[\mu} e_{\nu]}{\!}^{b]}
	+e^{\nu[a}e^{b]\rho}e_{\mu c}\partial_\rho e_\nu{\!}^c+\text{Fermions}$.},
\begin{align}
\partial_\mu \e +\frac{1}{4}\Gamma _{ab} \omega_\mu{}^{ab} \e+\frac{d_1}{4}\Gamma_{ab} F^{ab} \Gamma_\mu \e=0,
\end{align}
where $\e$ is a Dirac spinor, and $d_1=\sqrt{c_1}$ with $c_1 = \pm 1$. In the following sections, we will consider the ``electric'' set-up, and therefore we take $d_1=c_1=1$. For the ``magentic'' set-up one takes $c_1=-1, d_1= \mathrm{i} $ \cite{Dunajski:2006vs,Dunajski:2010uv}. In appendix \ref{app:conventions}, we derive the latter by using Euclidean superconformal tensor calculus.

\subsection{Euclidean Kerr-Newman solution saturating the BPS bound}

The Lorentzian Kerr-Newman solution in Boyer-Lindquist coordinates take the form
\bea
\label{knboyer}
ds^2 &=& - {r^2 + a^2 \cos^2\theta - 2 m r + q^2\over 
r^2+ a^2 \cos^2\theta} \, dt^2 + \frac{r^2+ a^2 \cos^2\theta}{r^2 + a^2 - 2 m r + q^2} dr^2 + (r^2+ a^2 \cos^2\theta) d\theta^2 \nonumber\\
&&+\frac{(r^2+ a^2 \cos^2\theta) (r^2 + a^2) + (2 m r - q^2) a^2 \sin^2 \theta}{r^2+ a^2 \cos^2\theta} \sin^2\theta d\phi^2 \nonumber \\
&&+\frac{2(q^2 - 2 m r)a}{r^2+ a^2 \cos^2\theta} \sin^2 \theta dt d\phi \, ,\nonumber \\
B_\mu dx^\mu &=& -{q \, r\over r^2 + a^2 \cos^2 \theta} \left[ dt - a \sin^2\theta
d\phi \right].
\eea
We call the Lorentzian vector field $B_\mu$. 
For these fields, Einstein equations hold in the following form\footnote{We use the following conventions for the Riemann tensor \cite{Freedman:2012zz}: $R_{\mu \nu}{}^{\rho}{}_{\sigma} = \partial_\mu \Gamma^\rho_{\nu \sigma}- \partial_\nu \Gamma^{\rho}_{\mu \sigma} + \Gamma^\rho_{\mu \tau} \Gamma^\tau_{\nu \sigma} - \Gamma^\rho_{\nu \tau} \Gamma^\tau_{\mu \sigma} $. The Ricci tensor is: $R_{\mu \nu} = R_{\mu}{}^{\sigma}{}_{\nu\sigma}$.}:
\begin{equation}
  R_{\alpha \beta} = 2 \big( \cF_{\alpha}{}^{\gamma} \,  \cF_{\beta\gamma}  -  \coeff{1}{4}   \delta_{\alpha\beta}    \cF^{\gamma\delta} \cF_{\gamma\delta} \big),
\label{EinVsimp2}
\end{equation}
where $\cF = dB$. Note the overall positive sign on the RHS; also note that this solution carries  an \emph{electric} charge. The (outer) horizon in this geometry is located at
\be
r_+ = m + \sqrt{m^2 -q^2 -a^2}.
\ee
The inverse temperature and angular velocity at the horizon are given as
\be
\beta = 4 \pi \frac{(r_+^2 + a^2)r_+}{r_+^2 -a^2 -q^2}, \qquad \beta \Omega = \beta \frac{a}{r_+^2 +a^2} =  4 \pi \frac{a r_+}{r_+^2 -a^2 -q^2}.
\ee
The ADM mass of the solution using eq.~(14) of \cite{Emparan:2008eg} is $M_{ADM} = \frac{m}{G_N}$ where $G_N$ is the four-dimensional Newton's constant. A standard definition of the electric charge in four-dimensional Einstein-Maxwell theory is $Q_E = \frac{1}{4 \pi G} \int_{S^2_\infty} \cF_{rt}~d \Omega$. This gives $Q_E = \frac{q}{G_N}$.  With these normalisations, the BPS relation in minimal ${\cal N}=2$ supergravity is $M_{ADM} = Q_E$, which implies $m=q$. This condition also coincides with the integrability condition of the Killing spinor equation \cite{Iliesiu:2021are}. However, as emphasised in \cite{Iliesiu:2021are, Hristov:2022pmo}, it does not correspond to the saturation of the extremality bound, which, if imposed on top $m=q$, would also require to set the angular momentum parameter $a$ to zero.

Since we are interested in Euclidean solutions, we do the following analytic continuation
\bea
&& t = - \mathrm{i} \, \tau,  \\ 
&& a = \mathrm{i} \,  b, \\
&& B =  \mathrm{i} \, A_E.   
\eea
and set the mass parameter $m$ equal to the charge parameter $q$. 
The new fields are:
\bea
\label{knEboyer}
ds^2 &=&  {r^2 - b^2 \cos^2\theta - 2 q r + q^2\over 
r^2- b^2 \cos^2\theta} \, d\tau^2 + \frac{r^2 - b^2 \cos^2\theta}{r^2 - b^2 - 2 q r + q^2} dr^2 + (r^2 - b^2 \cos^2\theta) d\theta^2 \nonumber\\
&&+\frac{(r^2 - b^2 \cos^2\theta) (r^2 - b^2) - (2 q r - q^2) b^2 \sin^2 \theta}{r^2-  b^2 \cos^2\theta} \sin^2\theta d\phi^2 \nonumber \\
&&+\frac{2(q^2 - 2 q r)b}{r^2 - b^2 \cos^2\theta} \sin^2 \theta d\tau d\phi \, , \\
\label{knEboyer2}
A_E &=&  {q \, r\over r^2 - b^2 \cos^2 \theta} \left[ d\tau + b \sin^2\theta
d\phi \right]. 
\eea
This configuration solves,
\begin{equation}
  R_{\alpha \beta} = - 2 \big( F_{\alpha}{}^{\gamma} \,  F_{\beta\gamma}  -  \coeff{1}{4}   \delta_{\alpha\beta}    F^{\gamma\delta} F_{\gamma\delta} \big),
\label{EinVsimp2_E}
\end{equation}
where $F= dA_E$. Note the overall minus sign on the RHS.

In the Euclidean setting, the `horizon' corresponds to 
\be
r_+ = q + b.
\ee
To check the regularity at this surface, we define 
\be
 \tilde r = \sqrt{r - r_+}, \qquad \tilde \phi = \phi -  \frac{b}{r_+^2 - b^2} \tau.
\ee  
Near the horizon, the metric takes the form (see, e.g., \cite{H:2023qko}) 
\bea
ds^2 & \simeq& \left[2 \frac{(r_+^2 - b^2 \cos^2 \theta)}{b} \right] \left\{d \tilde r^2 + \frac{b^2}{(r_+^2 -b^2)^2}  \tilde r^2 d \tau^2 \right\} \nn \\ &+& (r_+^2 -b^2 \cos^2 \theta)  d \theta^2 + \frac{\sin^2 \theta (r_+^2 - b^2)^2}{r_+^2 -b^2 \cos^2 \theta}d \tilde \phi^2. \label{smooth_metric} 
\eea
In the $(\tilde r, \tau)$ plane, metric \eqref{smooth_metric} is smooth provided $\frac{b}{(r_+^2 -b^2)} \tau $ is an angular coordinate with periodicity $2\pi$. That is, 
\be
(\tau, \tilde \phi)  \equiv \left( \tau + \beta , \tilde \phi \right),
\ee
where $\beta = 2 \pi  b^{-1} (r_+^2 - b^2)$.  This implies,
\be
(\tau, \phi)  \equiv \left( \tau + \beta , \phi + 2 \pi \right).  \label{identifications_4d_1}
\ee
Metric \eqref{smooth_metric} is also smooth at $\cos\theta = \pm 1$ provided
\be
(\tau, \phi)  \equiv \left( \tau, \phi + 2 \pi \right).  \label{identifications_4d_2}
\ee
For the vector field, a constant additive shift ensures that the integral of $A_E$ along the contractible cycle $\tau$ vanishes at $r=r_+$.

In summary, Euclidean Kerr-Newman solution saturating the BPS bound $m=q$ is a perfectly smooth manifold with identifications \eqref{identifications_4d_1}--\eqref{identifications_4d_2}.
Our aim below is to show that this configuration admits smooth normalisable Killing spinors. 
Since $\beta$ is finite, we call these solutions finite temperature solutions. As Lorentzian metrics,  these are not real and smooth. In the Lorentzian signature, smoothness requires additional conditions on the angular momentum parameter $a$ \cite{Iliesiu:2021are, Hristov:2022pmo,Cabo-Bizet:2018ehj}; in the present context it needs to be set to zero.

\subsection{Euclidean Israel-Wilson-Perjes (IWP) metrics}
  A class of supersymmetric Euclidean electrovac solution are the Israel-Wilson-Perjes (IWP) metrics. They take the form \cite{Perjes:1971gv, Israel:1972vx,  Whitt:1984wk, Yuille:1987vw, Dunajski:2006vs}:
\begin{equation}
ds^2_4 = (V_+ \,  V_-)^{-1}(d\tau+  A)^2 + (V_+ \,  V_-) (dy_1^2 +dy_2^2 + dy_3^2) \,,
\end{equation}
where
\be
\star_3 d A = V_-  d V_+  -V_+  d V_-\, ,
\label{Adefn}
\ee
where $\star_3$ is the Hodge star with respect to three-dimensional flat space and the functions $V_\pm$  are required to be harmonic on the $\IR^3$ base. 
 Introducing the frames:
\begin{equation}
 e^1~=~  (V_+ \,  V_-)^{-{1\over 2}}\, (d\tau +  A) \,,
\qquad   e^{a+1} ~=~  (V_+ \,  V_-)^{1\over 2}\, dy^a \,, \quad a=1,2,3 \,,
\label{IWframes}
\end{equation}
the background Maxwell field is given by \cite{Dunajski:2006vs, Bena:2009fi}:
\begin{eqnarray}
\widetilde {\cF} 
&=&  - \coeff{1}{2} \big[ \partial_a \big(V_+^{-1} - V_-^{-1} \big) \big] \,   e^1 \wedge    e^{a+1} ~-~
 \,\coeff{1}{4} \epsilon_{abc}  \big[ \partial_a \big(V_+^{-1} + V_-^{-1} \big) \big]    e^{b+1} \wedge    e^{c+1}    \,.
\label{IWFform}
\end{eqnarray}
This background then satisfies Euclidean Einstein-Maxwell equations:
\begin{equation}
  R_{\alpha \beta} =  2 \big( \widetilde {\cF}_{\alpha}{}^{\gamma} \,  \widetilde {\cF}_{\beta\gamma}  -  \coeff{1}{4}   \delta_{\alpha\beta}    \widetilde {\cF}^{\gamma\delta} \widetilde {\cF}_{\gamma\delta} \big).
\label{EinVsimp}
\end{equation}
We note that when $V_+ =  V_- $ the metrics reduce to Majumdar-Papapetrou metrics with purely \emph{magnetic} field strength.

\subsection{Kerr-Newman in IWP form}
As we will see shortly, it is easier (and more useful as it has wider applicability) to workout Killing spinors for the general IWP metrics. In order to show that the 
Euclidean Kerr-Newman solution saturating the BPS bound is supersymmetric, we first express it in the IWP form. To do so, we relate coordinates $(r, \theta, \phi)$ appearing in the Boyer-Lindquist form to the cartesian  $\IR^3$ base coordinates $(y_1, y_2, y_3)$ as,
\begin{align}
y_1 &= \sqrt{(r -q )^2 - b^2} \, \sin \theta \, \cos \phi, &
y_2 &= \sqrt{(r -q )^2 - b^2} \, \sin \theta \,  \sin \phi, &
y_3 &= (r -q ) \, \cos \theta.
\end{align}
From these definitions it follows that $(r, \theta, \phi)$ are some variant of the ``oblate spheroidal'' coordinates, such that
\be
\frac{y_1^2+ y_2^2}{(r-q)^2 - b^2} + \frac{y_3^2 }{(r-q)^2} = 1. 
\ee
The variable $r$ is a shifted radial coordinate. 
Take $r_+$ to be the distance from the point $ (0,0,b)$
\be
r_+   = \sqrt{y_1^2 + y_2^2 + (y_3 - b)^2}
=  r - q - b \cos \theta, 
\ee
and $r_-$ to be the distance from the point $ (0,0,-b)$
\be
r_- = \sqrt{y_1^2 + y_2^2 + (y_3 + b)^2}
=  r - q + b \cos \theta. 
\ee
The choice, 
\begin{align}
V_+ &= 1 + \frac{q}{r_+}, &
V_- &= 1 + \frac{q}{r_-},
\end{align}
gives the Euclidean Kerr-Newman metric eq.~\eqref{knEboyer}.  The electric form field from eq.~\eqref{knEboyer2} is the four-dimensional Hodge dual of eq.~\eqref{IWFform}:
\begin{align}
F &= dA_E = \star_4 \,  \widetilde {\cF} &\\
&= -\coeff{1}{4} \epsilon_{abc}   \big[ \partial_a \big(V_+^{-1} - V_-^{-1} \big) \big] \,   e^{b+1} \wedge    e^{c+1} ~-~
 \, \coeff{1}{2} \big[ \partial_a \big(V_+^{-1} + V_-^{-1} \big) \big]    e^{1} \wedge    e^{a+1} . &
\label{IWFformE}
\end{align}
To see this more explicitly, note that, 
\begin{align}
e^2 \wedge e^3 &= V_+ V_- dy_1 \wedge dy_2, &
e^2 \wedge e^4 &= V_+ V_- dy_1 \wedge dy_3, &
e^3 \wedge e^4 &= V_+ V_- dy_2 \wedge dy_3. 
\end{align}
On the three-dimensional base space   $\epsilon_{y_1 y_2 y_3} = 1$; in Boyer-Lindquist coordinates 
\be
\epsilon_{r \theta \phi} =  ((r-q)^2 - b^2 \cos^2 \theta)\sin \theta.
\ee
Integration of eq.~\eqref{Adefn} gives\footnote{It is easiest to integrate equation \eqref{Adefn} in $(r, \theta, \phi)$ coordinates. For the rest of this subsection, one can think of $A_\phi$ as a function of $(y_1, y_2, y_3)$ whose form in the $(r, \theta, \phi)$ coordinates is \eqref{A_phi_form}. Explicitly,
\be
A_\phi = q^2 \left( \frac{y_1^2 + y_2^2 + (y_3 - b + r_+)(y_3 + b- r_-)}{2b r_+ r_-}\right) + \left(q + \frac{q^2}{2b} \right) \left(\frac{y_3 - b}{r_+} -  \frac{y_3 + b}{r_-} \right).
\ee }
\be
A = A_\phi d\phi = \frac{q b (q-2 r)}{(r-q)^2-b^2 \cos ^2\theta
   } \sin^2 \theta d\phi. \label{A_phi_form}
\ee
Using, 
$\phi = \tan^{-1} \frac{y_2}{y_1}$, 
we can convert the three-dimensional one form $A = A_\phi d \phi$ in the $y_1, y_2, y_3$ coordinates as
\be
A = \frac{A_\phi}{y_1^2 + y_2^2} (y_1 d y_2 - y_2 d y_1). 
\ee
Then,
\bea
e^1 \wedge e^2 &=& d \tau \wedge d y_1 - \frac{A_\phi}{y_1^2 + y_2^2} y_1 dy_1 \wedge d y_2,  \\
e^1 \wedge e^3 &=&  d \tau \wedge d y_2 - \frac{A_\phi}{y_1^2 + y_2^2} y_2 dy_1 \wedge d y_2, \\
e^1 \wedge e^4 &=&d \tau \wedge d y_3 + \frac{A_\phi}{y_1^2 + y_2^2} (y_1 d y_2 \wedge dy_3 - y_2 d y_1 \wedge dy_3).
\eea
All this implies, 
\bea
F_{\tau a} &=&    - \coeff{1}{2} \partial_a \big(V_+^{-1} + V_-^{-1}\big),
\eea
and
\bea
F_{y_1 y_2} &=& - \coeff{1}{2} V_+ V_- \partial_{y_3}  \big(V_+^{-1} - V_-^{-1}\big)  - \frac{A_\phi}{y_1^2 + y_2^2}   (y_1  F_{\tau y_1} + y_2      F_{\tau y_2} ), \\
F_{y_1 y_3} &=&  + \coeff{1}{2} V_+ V_- \partial_{y_2}  \big(V_+^{-1} - V_-^{-1}\big)  - \frac{A_\phi}{y_1^2 + y_2^2}   y_2 F_{\tau y_3},\\
F_{y_2 y_3} &=& - \coeff{1}{2} V_+ V_- \partial_{y_1}  \big(V_+^{-1} - V_-^{-1}\big)  + \frac{A_\phi}{y_1^2 + y_2^2}   y_1 F_{\tau y_3}.
\eea
Computing these components, and doing the coordinate transformation, we get $F = dA_E$ for the 
Euclidean Kerr-Newman solution in the Euclidean Boyer-Lindquist form.  
\subsection{Killing spinors}

For the electric form of the IWP solutions, the field strength in the frame basis  \eqref{IWframes} takes the form \eqref{IWFformE}.
The combination $ F_{ab} \Gamma^{ab}$ is, 
\be
 F_{ab} \Gamma^{ab} = \left(\begin{array}{cc}
\frac{2 \mathrm{i} \vec{\tau} \, \cdot \, \vec{\nabla} V_+} {V_+^2} & 0 \\
0 & - \frac{2 \mathrm{i} \vec{\tau} \, \cdot \, \vec{\nabla} V_-} {V_-^2} 
\end{array}\right),
\ee
where the gamma matrix conventions are given in appendix \ref{app:gamma}; $\vec{\tau}\,$ are the Pauli matrices. For the spin connection combination $\omega_{ab} \Gamma^{ab}$ we have, 
\be
\omega_{ab} \Gamma^{ab} =
\frac{-2\mathrm{i}}{(V_+V_-)^{1/2}}  \left(\begin{array}{cc}
\frac{\vec{\t} \, \cdot
 \,  \vec{ \nabla}V_+ }{V_+} e^1+ \frac{({\vec \t} \, \times \, \vec{\nabla} V_+) }{V_+} \cdot \vec{e}  & 0 \\
0 &- \frac{\vec{\t} \, \cdot
 \,  \vec{ \nabla}V_-}{V_-} e^1+ \frac{({\vec \t} \, \times \, \vec{\nabla} V_-) }{V_-} \cdot \vec{e}
\end{array}\right),
\ee
where $\vec{e} = \{e_2, e_3, e_4\}.$ Now, we can easily solve for the Killing spinors
\begin{align}
\partial_\mu \e +\frac{1}{4}\Gamma _{ab} \omega_\mu{}^{ab} \e+\frac{1}{4}\Gamma_{ab} F^{ab} \Gamma_\mu \e=0.
\end{align}
 They take the form, 
\be
\e =
\left( \begin{array}{c}
- V_+^{-1/2} \e_0 \\
 V_-^{-1/2} \e_0
\end{array}
\right) \,, \label{eq:spinor_4d}
\ee
where $\e_0$ is a constant two-component complex spinor: $\partial_\mu
\e_0 = 0$. For the magnetic IWP solutions, Killing spinors were written in \cite{Dunajski:2006vs}. Under $2 \pi$ rotation in the $(y_1,y_2)$ plane, $\e$ goes to $\mathrm{exp}\left[ (2 \pi) \cdot 2 \cdot  \frac{1}{4} \Gamma^{23} \right] \e= -\e$. Thus,  Killing spinors \eqref{eq:spinor_4d} satisfy the expected periodicity properties under the identifications \eqref{identifications_4d_1}--\eqref{identifications_4d_2}.   This is so because the Killing spinors have no dependence on the $\tau$ coordinate. Therefore, as far as the action of identification \eqref{identifications_4d_1}
on the Killing spinors in concerned, it simply corresponds to $2 \pi$ rotation in the $(y_1,y_2)$ plane.  Since, $V_\pm$ go to 1 at infinity, the Killing spinors are normalisable, unlike in the near horizon geometry where they are non-normalisable as we approach the boundary of AdS$_2$ \cite{Sen:2023dps}. This is one of the attractive features of the recent developments in computing the index using path integral over supergravity fields in the full asymptotically flat geometry.

\section{Euclidean Cvetic-Youm solution}
\label{sec:5d}
 Anupam, Chowdhury, and Sen (ACS) \cite{Anupam:2023yns} conjectured that Euclidean five-dimensional Cvetic-Youm solution saturating the BPS bound should also admit Killing spinors. The aim of this section is to establish this. We follow the same logic as above. In order to show that the Euclidean Cvetic-Youm solution saturating the BPS bound is supersymmetric, we express it in the Gibbons-Hawking form. It is easier, and more useful as it has wider applicability, to workout Killing spinors for a general class of Gibbons-Hawking metrics. The notation used by ACS is somewhat unwieldy. One of our key technical observation is that the ACS metric can be conveniently written in the Chong, Cvetic, Lu, Pope (CCLP) form \cite{Chong:2005hr}. This observation makes the calculations manageable. In appendix \ref{app:ACS}, we present a detailed comparison between the metric used below and the metric given in the ACS paper.

\subsection{Euclidean five-dimensional minimal supergravity}
In this section, we review the equations of motion and the Killing spinor equation for five dimensional Euclidean minimal supergravity. The bosonic sector again is the five dimensional Einstein-Maxwell theory with the equations of motion,
\be
R_{\mu \nu}=- \frac{1}{2} \left(F_{\mu\rho}F_{\nu}{}^{\rho}-\frac{1}{6}F_{\rho \sigma} F^{\rho \sigma} g_{\mu\nu} \right),
\label{einstein-eq-5d}
\ee
and
\be
d \left(\star_5 F - \frac{1}{\sqrt{3}} F \wedge A \right) = 0,
\label{vector-eq}
\ee
where the vector equation of motion contains a contribution from the five-dimensional Chern-Simons term. The Killing spinor equation for the theory reads,
\be
\partial_\mu \e + \frac{1}{4} \omega_{\mu}{}_{ab} \Gamma^{ab} \e - \frac{1}{8\sqrt{3}} \left(\Gamma_\mu \Gamma_{ab} F^{ab} - 6  F_{\mu \nu} \Gamma^{\nu} \right) \e =0. \label{Killing-spinor-eq-5d}
\ee
In \cite{Sabra:2016abd}, the Lagrangian and Killing spinor equation for this theory were constructed by oxidisation from the four dimensional Euclidean $\mathcal{N}=2$ theory. We can confirm the consistency of the Killing spinor equations and the equations of motion, by performing the integrability analysis.\footnote{Our conventions for the equations of motion are different from that of \cite{Sabra:2016abd} and are fixed such that upon analytic continuation they match the conventions of \cite{Chong:2005hr}. The integrability analysis checks that our conventions are self consistent.}

We can take an ansatz for the Killing spinor equation,
\be
\partial_\mu\e+\frac{1}{4}\omega_\mu{}^{ab}\Gamma_{ab}\e+d_2 \Gamma_{ab} F^{ab} \Gamma_\mu\e+d_3 \Gamma_\mu\Gamma_{ab} F^{ab} \e=0.
\ee
Demanding that the integrability condition is consistent with the equations of motion above we get,
\begin{align}
 (3 d_2^2 + 2 d_2 d_3 + 3 d_3^2) &= 
\frac{1}{32},\hspace{1cm}
(d_2 + d_3)^2 = \frac{1}{192}, \hspace{1cm} \frac{d_2d_3}{(d_2 + d_3)} =\frac{\sqrt{3}}{32}.
\end{align}
One can check that $d_2=-\frac{\sqrt{3}}{16},d_3=\frac{1}{16\sqrt{3}}$ is a solution to these equations. Thus the Killing spinor equation 
\be
\partial_\mu\e+\frac{1}{4}\omega_\mu^{ab}\Gamma_{ab}\e-\frac{\sqrt{3}}{16} \Gamma_{ab} F^{ab} \Gamma_\mu\e+\frac{1}{16\sqrt{3}} \Gamma_\mu \Gamma_{ab} F^{ab} \e=0,
\ee
is consistent with the Einstein-Maxwell equations of motion \eqref{einstein-eq-5d} and \eqref{vector-eq}. By using properties of gamma matrices, one can arrive at \eqref{Killing-spinor-eq-5d}.

It is desirable to obtain these equation from the superconformal approach. However, to the best of our knowledge, Euclidean five-dimensional conformal supergravity has not been constructed analogous to the four-dimensional case. However, from Wick rotation, superconformal multiplets and their transformations have been proposed in \cite{Ciceri:2023mjl}. For our analysis, a consistent set of equations of motion and the Killing spinor equation given above suffice.

\subsection{Euclidean Cvetic-Youm solution saturating the BPS bound}

The Lorentzian Cvetic-Youm solution \cite{Cvetic:1996xz, Cvetic:1997uw} can be conveniently written in the Boyer-Lindquist type coordinates $x^\mu= (t, r, \theta,
\phi, \psi)$ as follows \cite{Chong:2005hr}\footnote{This form is obtained upon setting  $g=0$ in \cite{Chong:2005hr}.}.  Define,
\bea
\Delta_r &=& \fft{(r^2+a^2)(r^2+b^2)  + q^2 +2ab q}{r^2} - 2m 
\,,\\
\rho^2 &=& r^2 + a^2 \cos^2\theta + b^2 \sin^2\theta\, ,
\eea
then,
\bea
\label{metric_5d_first}
g_{tt}\!\!\!&=&\!\!\! -1 + 
         \fft{(2m \rho^2 -q^2)}{
         \rho^4}\,,\\
g_{rr}\!\!\! &=&\!\!\! \fft{\rho^2}{\Delta_r}\,, \\
 g_{\theta \theta}\!\!\! &=&\!\!\! \rho^2\,,\\
g_{t \psi}\!\!\! &=&\!\!\! -\fft{[b (2m \rho^2-q^2) + 
     a q\rho^2]\cos^2\theta }{\rho^4}\,,\\
g_{t \phi}\!\!\! &=&\!\!\! -\fft{ [a (2m \rho^2-q^2) + 
     b q\rho^2 ]\sin^2\theta}{\rho^4 }\,,\\
g_{\psi\psi}\!\!\! &=&\!\!\! (r^2+b^2)\cos^2\theta + 
                \fft{ b[b(2m\rho^2 -q^2) + 2 aq \rho^2]\cos^4\theta}{
              \rho^4}\,,\\
g_{\phi\phi}\!\!\! &=&\!\!\!  (r^2+a^2)\sin^2\theta  + 
                \fft{ a[a(2m\rho^2 -q^2) + 2 bq \rho^2]\sin^4\theta}{
              \rho^4}\,,\\
g_{\phi\psi}\!\!\!&=&\!\!\!
\fft{[ ab(2m \rho^2 -q^2) + (a^2+b^2) q \rho^2]\, \sin^2\theta\,
             \cos^2\theta}{\rho^4 }\,.
\label{metric_5d_last}
\eea
The vector field supporting the Lorentzian solution is,
\be
A = \frac{\sqrt{3}q}{\rho^2}\left(dt - a \sin^2 \theta d\phi - b \cos^2\theta d\psi \right).\label{vector_5d_L}
\ee

The largest positive root of $\Delta_r=0$ is the location of the outer Killing horizon at $r=r_+$. At the horizon, the Killing vector 
\be
\ell = \fft{\partial}{\partial t} + \Omega_\phi\, \fft{\partial}{\partial \phi} 
    + \Omega_\psi \, \fft{\partial}{\partial \psi},
\ee
becomes null, where he angular velocities  $\Omega_\psi$ and $\Omega_\phi$ are
\begin{align}
\Omega_\phi &= \fft{a(r_+^2+ b^2) + b q}{
               (r_+^2+a^2)(r_+^2+b^2)  + ab q}, & \label{Omega_phi_psi} 
\Omega_\psi &= \fft{b(r_+^2+ a^2) + a q}{
               (r_+^2+a^2)(r_+^2+b^2)  + ab q}.
\end{align}
The surface gravity at the outer Killing horizon evaluates to,
\be
\kappa = \fft{r_+^4 -(ab + q)^2}{
         r_+\, [(r_+^2+a^2)(r_+^2+b^2) + abq]}. \label{kappa}
\ee

The ADM mass of the solution using eq.~(14) of \cite{Emparan:2008eg} is
$M_{ADM} = \frac{3 \pi m}{4G_N}$ where now $G_N$ is the five-dimensional Newton's constant. A standard definition of electric charge in minimal five-dimensional supergravity is $Q_E = \frac{1}{16 \pi G_N} \int_{S^3_\infty} (\star_5 F - F \wedge A/\sqrt{3})~d \Omega$. This gives $Q_E = \frac{\sqrt{3} \pi q}{4 G_N}$.  With these normalisations, the BPS relation in minimal five-dimensional supergravity  is $M_{ADM} = \sqrt{3} \, Q_E$ \cite{Chong:2005hr}, which implies $m=q$. However, as emphasised in \cite{Cabo-Bizet:2018ehj} (in the AdS context), it does not correspond to the saturation of the extremality bound, which, if imposed on top $m=q$, would also put constraints on the angular momentum parameters.

We are interested in Euclidean solutions. Euclidean Cvetic-Youm solution saturating the BPS bound can be obtained by setting $m=q$ and doing the analytic continuation
\begin{align}
& t = -\mathrm{i} \, \tau, & \label{analytic_continuation_1} \\
& a = \mathrm{i} \, \bar{a}, & \label{analytic_continuation_2} \\
&  b = - \mathrm{i} \, \bar{b}.  \label{analytic_continuation_3}
\end{align}
The resulting metric is real. It can be written in the form,
\be
ds^2 = f^2 (d \tau + \omega_\phi d\phi + \omega_\psi d\psi)^2 + f^{-1} ds_4^2
\ee
where $ds_4^2$ is the metric on a four-dimensional base. The function $f$  takes the form, 
\be
f = 1- \frac{q}{\rho^2}, 
\ee
with 
\be
\rho^2 = r^2 -\bar{a}^2 \cos^2 \theta - \bar{b}^2 \sin^2 \theta,
\ee
and
\begin{align}
\omega_\psi &=   q \cos^2 \theta \frac{(1+ f)\bar{b}  - \bar{a} }{f^2 \rho^2}, &
\omega_\phi &=  -q \sin^2 \theta \frac{(1+ f)\bar{a}  - \bar{b} }{f^2 \rho^2}.
\end{align}
The four-dimensional  base in $(r, \theta, \phi, \psi)$ coordinates  is,
\begin{align}
g^{4d}_{rr} &= \frac{f \rho^2}{\Delta_r} \label{base_4d_1}, &
g^{4d}_{\theta \theta} &= f \rho^2 , &
g^{4d}_{\phi\phi} &=  f (g_{\phi \phi} - f^2 \omega_\phi^2), & \\
g^{4d}_{\psi\psi} &= f (g_{\psi \psi} - f^2 \omega_\psi^2), &
g^{4d}_{\phi\psi} &= f (g_{\phi \psi} - f^2 \omega_\phi \omega_\psi). \label{base_4d_5}
\end{align}
The base is Ricci-flat. This is the second hint (the first being it saturates the BPS bound) that the 
Euclidean Cvetic-Youm solution saturating the BPS bound is supersymmetric.

To establish the smoothness properties of the solution, we observe that the location\footnote{Without any loss of generality we restrict ourselves to $\bar{a} \ge \bar{b}$.} of the horizon in the Euclidean metric is,
\be
r_+ = \frac{1}{2} \left( \bar{a} - \bar{b} + \sqrt{(\bar{a} + \bar{b})^2 + 4 q}\right).
\ee
Proceeding along the lines of the case of the Kerr-Newman solution discussed in the previous section we define,
\begin{align}
 \tilde r &= \sqrt{r - r_+}, &
  \tilde \phi &= \phi -  \Omega_\phi \tau, &
   \tilde \psi &= \psi -  \Omega_\psi \tau,&
\end{align}
where
\begin{align}
\Omega_\phi &= \fft{\bar{a}(r_+^2 - \bar{b}^2) -  \bar{b} q}{
               (r_+^2 - \bar{a}^2)(r_+^2 -\bar{b}^2)  + \bar{a}\bar{b} q}, &
\Omega_\psi &= \fft{-b(r_+^2-\bar{a}^2) +  \bar{a} q}{
               (r_+^2- \bar{a}^2)(r_+^2-\bar{b}^2)  + \bar{a}\bar{b} q}.
\end{align}
Expressions for $\Omega_\phi$ and $\Omega_\psi$ are obtained via analytic continuation \eqref{analytic_continuation_2}--\eqref{analytic_continuation_3} of \eqref{Omega_phi_psi}, respectively. 

Near the horizon,  the $(\tilde r, \tau)$ part of the metric takes the form,
\bea
ds^2 & \simeq& \left[2 r_+ \frac{(r_+^2 - \bar{a}^2 \cos^2 \theta - \bar{b}^2 \sin^2 \theta)}{(\bar a - \bar b) \sqrt{(\bar a + \bar b)^2+4q}} \right] \left\{d \tilde r^2 +   \kappa^2  \tilde r^2 d \tau^2 \right\}, \label{smooth_metric_5d} 
\eea
where
\be
\kappa = \fft{r_+^4 -(\bar{a}\bar{b} + q)^2}{
        r_+\, [ (r_+^2- \bar{a}^2)(r_+^2-\bar{b}^2)  + \bar{a}\bar{b} q]}.
\ee
Metric \eqref{smooth_metric_5d} is smooth provided $(\kappa \tau)$ is an angular coordinate with periodicity $2\pi$. That is, 
\be
(\tau, \tilde \phi, \tilde \psi)  \equiv \left( \tau + \beta , \tilde \phi, \tilde \psi \right),
\ee
where $\beta = 2 \pi \kappa^{-1}$. This implies,
\be
(\tau,  \phi, \psi)  \equiv \left( \tau +\beta,  \phi + \beta \Omega_\phi,  \psi +  \beta \Omega_\psi \right),
 \label{identifications_5d}
\ee
We note that  $\beta(\Omega_\phi + \Omega_\psi) = 2 \pi$.\footnote{Having set $m=q$,  we do not have any option left in the choice of $\beta(\Omega_\phi + \Omega_\psi)$. We thank Ashoke Sen and Sameer Murthy on discussions on this point.}
The gravitational path integral that computes the index  in five-dimensions instructs us to set angular velocities such that $\beta(\Omega_\phi + \Omega_\psi) = 2 \pi$. 

We also have the identifications
\be
(\tau,  \phi, \psi)  \equiv (\tau,  \phi + 2 \pi, \psi)  \equiv (\tau,  \phi , \psi + 2 \pi).
 \label{identifications_5d_2}
\ee
To see that identifications \eqref{identifications_5d_2} lead to smooth metric we need to analyse the 
$(\theta, \tilde \phi, \tilde \psi)$ part of the metric\footnote{The cross terms $g_{\tau \tilde \phi}$ and  $g_{\tau \tilde \psi}$ vanish in the $\tilde r \to 0$ limit.} at $r=r_+$. The full $(\theta, \tilde \phi, \tilde \psi)$ part of the metric is not illuminating. Since we are interested in examining the smoothness of the metric, we only need to analyse the metric 
near the north and the south poles $\theta =0$ and $\theta = \pi/2$, respectively. Near $\theta = 0$, it takes the form,
\bea
ds^2 = c_{\theta \theta} d \theta^2  + c_{\tilde \phi \tilde \phi} d  \tilde \phi^2 + 2 c_{\tilde \phi \tilde \psi} d \tilde \phi d \tilde \psi + c_{\tilde \psi \tilde \psi} d \tilde \psi^2,
\eea
where
\begin{align}
&c_{\theta \theta} = r_+^2 - \bar{a}^2 + \mathcal{O}(\theta^2), \\ 
&c_{\tilde\phi \tilde\phi} = (r_+^2 - \bar{a}^2) \theta^2  + \mathcal{O}(\theta^4) ,\\
&c_{\tilde\phi \tilde\psi} = -\frac{r_+ + \bar{b}}{(r_+ + \bar{a})^2} \left( \left((\bar{a} - \bar{b})^2 +  \bar{a} \bar{b} \right)  r_+ + \bar{a}((\bar{a} - \bar{b})^2 + \bar{b}^2) \right)\theta^2+ \mathcal{O}(\theta^4) , \\
&c_{\tilde\psi \tilde\psi} = \frac{(r_+ + \bar{a} - \bar{b})^2(r_+ + \bar{b})^2}{(r_+ +\bar{a})^2} + \mathcal{O}(\theta^2).
\end{align}
The $\tilde\psi$ circle has finite size. We see that if we shift $\tilde {\phi}  $ to $ \tilde{\tilde {\phi}} $ as,
\be
\tilde {\phi} = \tilde{\tilde {\phi}}   + \frac{(r_+ + \bar{b})}{(r_+ + \bar{a})^2(r_+^2 - \bar{a}^2)} \left( \left((\bar{a} - \bar{b})^2 +  \bar{a} \bar{b} \right)  r_+ + \bar{a}((\bar{a} - \bar{b})^2 + \bar{b}^2) \right) \tilde \psi,
\ee 
then the $\tilde{\tilde {\phi}} $ circle smoothly shrinks to zero size at $\theta =0$ for fixed $\tilde \psi$. The metric takes the form, 
\be
ds^2 = (r_+^2 - \bar{a}^2)(d \theta^2 + \theta^2 d \tilde{\tilde {\phi}}^2) + c_{\tilde\psi \tilde\psi} d \tilde \psi^2.
\ee
Thus, the correct identification is,
\be
(\tilde{\tilde {\phi}}, \tilde \psi) \equiv (\tilde{\tilde {\phi}} + 2 \pi, \tilde \psi)
\implies (\tilde {\phi}, \tilde \psi) \equiv (\tilde {\phi} + 2 \pi, \tilde \psi) 
\implies ( \phi,  \psi, \tau ) \equiv (\phi + 2 \pi, \psi, \tau).
\ee

To summarise, with the standard $2 \pi$ identification for the $\phi$ circle the metric is smooth at the north pole $\theta = 0$.

Near $\theta = \pi/2$ the analysis is very similar. The metric takes the form,
\bea
ds^2 = d_{\theta \theta} d \theta^2  + d_{\tilde \phi \tilde \phi} d  \tilde \phi^2 + 2 d_{\tilde \phi \tilde \psi} d \tilde \phi d \tilde \psi + d_{\tilde \psi \tilde \psi} d \tilde \psi^2,
\eea
where
\begin{align}
&d_{\theta \theta} = r_+^2 - \bar{b}^2 + \mathcal{O}((\theta-\pi/2)^2), \\ 
&d_{\tilde\psi \tilde\psi} = (r_+^2 - \bar{b}^2) (\theta-\pi/2)^2  + \mathcal{O}((\theta-\pi/2)^4) ,\\
&d_{\tilde\phi \tilde\psi} = -\frac{r_+ - \bar{a}}{(r_+ - \bar{b})^2} \left( \left((\bar{a} - \bar{b})^2 +  \bar{a} \bar{b} \right)  r_+ - \bar{b}((\bar{a} - \bar{b})^2 + \bar{a}^2) \right)(\theta-\pi/2)^2+ \mathcal{O}((\theta-\pi/2)^4) , \\
&d_{\tilde\phi \tilde\phi} = \frac{(r_+ + \bar{a} - \bar{b})^2(r_+ - \bar{a})^2}{(r_+ -\bar{b})^2} + \mathcal{O}(\theta^2).
\end{align}
Now, the $\tilde\phi$ circle has finite size. We see that if we shift $\tilde {\psi}  $ to $ \tilde{\tilde {\psi}} $ as,
\be
\tilde {\psi} = \tilde{\tilde {\psi}}   + \frac{r_+ - \bar{a}}{(r_+ - \bar{b})^2(r_+^2 - \bar{b}^2) } \left( \left((\bar{a} - \bar{b})^2 +  \bar{a} \bar{b} \right)  r_+ - \bar{b}((\bar{a} - \bar{b})^2 + \bar{a}^2) \right) \tilde \phi,
\ee 
then the $\tilde{\tilde {\psi}} $ circle smoothly shrinks to zero size at $\theta = \pi/2$ for fixed $\tilde \phi$. Thus, the correct identification is,
\be
(\tilde {\phi}, \tilde{\tilde \psi}) \equiv (\tilde {\phi} , \tilde{\tilde{ \psi}} + 2 \pi )
\implies (\tilde {\phi}, \tilde \psi) \equiv (\tilde {\phi} , \tilde \psi + 2 \pi) 
\implies ( \phi,  \psi, \tau ) \equiv (\phi , \psi+ 2 \pi, \tau).
\ee
We conclude that with the standard $2 \pi$ identification for the $\psi$ circle the metric is smooth at the south pole $\theta = \pi/2$.

 The vector field \eqref{vector_5d_L} upon analytic continuation takes the form,
\be
A_E = \mathrm{i} A  = \frac{\sqrt{3}q}{\rho^2}\left(d\tau + \bar {a} \sin^2 \theta d\phi - \bar{b} \cos^2\theta d\psi \right).
\ee
As for the Euclidean Kerr-Newman case discussed above,  a constant additive shift 
for the vector ensures that the integral of $A$ along the contractible cycle $\tau$ vanishes at $r=r_+$. For the Euclidean solution, the order of coordinates consistent with our conventions is $(\tau, r, \theta, \psi, \phi)$.

Since $\beta$ is finite, we call these solutions finite temperature Euclidean solutions. In this sense, these are non-extremal Euclidean solutions. As Euclidean metrics these are real and smooth. However, as Lorentzian metrics these are not real and smooth. In the Lorentzian signature, smoothness requires additional conditions on the angular momentum parameters \cite{Cabo-Bizet:2018ehj}. 

\subsection{Solution in Gibbons-Hawking form}
Our next key observation is that the four-dimensional base metric \eqref{base_4d_1}--\eqref{base_4d_5} can be written in the Gibbons-Hawking form. To do this, we proceed in steps. Let us first consider the special case of the metric when $\bar a = \bar b$. In this case, the (Lorentzian) black hole carries only left angular momentum. The Lorentzian metric is nothing but the supersymmetric BMPV black hole. In the case of the BMPV black hole, it is well know that the four-dimensional base in flat-space \cite{Gauntlett:2002nw}. The Euclidean continuation does not bring any new element. Thus, we expect the the four-dimensional base upon setting $\bar a = \bar b$ to be flat space. Indeed, it takes the form:
\be
ds^2_{4d} = \frac{r^2}{r^2 - q - \bar{b}^2} dr^2 + (r^2 - q - \bar{b}^2) ( d\theta^2 + \sin^2 \theta d\phi^2 + \cos \theta^2 d\psi^2),
\ee
Clearly, defining the radial variable to  $\sqrt{r^2 - q - \bar{b}^2}$, we immediately recognise this as flat space.

This suggests that in the general case, it is better to work with the combinations ($\bar a+\bar b$, $\bar a-\bar b$) in place of ($\bar a$, $\bar b$)  and ($\phi+ \psi$, $\psi - \phi$) instead of ($\phi$, $\psi$). We define, 
\begin{align}
\psi &= \frac{1}{2} \left( \varphi + \Psi \right), &
\phi &= \frac{1}{2} \left( \varphi - \Psi \right),
\end{align}
and
\begin{align}
\bar a &=  c +d , &
\bar b &=  c-d .
\end{align}
We next define, the new radial and angular coordinates $(R, \Theta)$ as,
\bea
R &=& \frac{1}{4} r^2 - \frac{1}{4} (d^2 + q), \\
\Theta &=& 2 \theta.
\eea
and further define, 
\bea
y_1 &=& \sqrt{\left(R-\frac{c^2}{4} \right)^2 - \frac{1}{4} c^2 (d^2 + q)} \,  \sin \Theta \cos \, \varphi, \\
y_2 &=&\sqrt{\left(R-\frac{c^2}{4} \right)^2 - \frac{1}{4} c^2 (d^2 + q)} \, \sin \Theta \sin \, \varphi , \\
y_3 &=& \left(R-\frac{c^2}{4} \right) \, \cos \Theta.
\eea

These transformations may look a bit mysterious at first, but  they can be appreciated as follows. The $(R,\Theta, \varphi)$ coordinates are ``oblate spheroidal'' coordinates constructed out of $(y_1, y_2, y_3)$. We have,
\be
\frac{y_1^2+ y_2^2}{\left(R-\frac{c^2}{4} \right)^2 - \frac{1}{4} c^2 (d^2 + q)} + \frac{y_3^2 }{\left(R-\frac{c^2}{4} \right)^2} = 1. 
\ee
As we will see shortly, $(y_1, y_2, y_3)$ or equivalently $(R, \Theta, \varphi)$ serve as  three-dimensional base coordinates with $\Psi$ as the Gibbons-Hawking fibre. The transformation from $(r, \theta)$ to $(R, \Theta)$ are similar to the one needed to write four-dimensional flat space from standard polar coordinates to Gibbons-Hawking coordinates. 

\subsubsection{4d base}

In coordinates $(\Psi, y_1, y_2, y_3)$  the four-dimensional base takes the standard Gibbons-Hawking form, 
\be
ds^2_{4} = V^{-1} (d \Psi  + A)^2 + V (dy_1^2 + dy_2^2 + dy_3^2),
\ee
where the function $V$ is harmonic. It is centred at two points with total charge unity. It takes the form, 
\be
V = \left(\frac{1}{2}-\frac{d}{2 \sqrt{d^2+q}} \right) \frac{1}{R_+} + \left(\frac{1}{2}+ \frac{d}{2 \sqrt{d^2+q}} \right) \frac{1}{R_-},
\ee
where $R_+$ to be the distance from the point $ \left(0,0,\frac{1}{2}c \sqrt{d^2 + q} \right)$ in $(y_1, y_2, y_3)$ space,
\bea
R_+   &=& \sqrt{y_1^2 + y_2^2 + \left(y_3 - \frac{1}{2}c \sqrt{d^2 + q} \right)^2}\\
&=&  R - \frac{1}{4}c^2 - \frac{1}{2} c \sqrt{d^2 + q} \cos \Theta, 
\eea
and $R_-$ to be the distance from the point $ \left(0,0,-\frac{1}{2}c \sqrt{d^2 + q} \right)$ in the $(y_1, y_2, y_3)$ space,
\bea
R_-   &=& \sqrt{y_1^2 + y_2^2 + \left(y_3 + \frac{1}{2}c \sqrt{d^2 + q} \right)^2}\\
&=&  R - \frac{1}{4}c^2 + \frac{1}{2} c \sqrt{d^2 + q} \cos \Theta.
\eea
The three dimensional one-form $A$ is simply,
\be
\star_3 dA = dV. 
\ee
The function $V$ represents the bound state of two Kaluza-Klein monopoles with net charge unity (also sometimes called in the literature two Taub-NUT bound state). This configuration indeed describes a four-dimensional asymptotically flat Euclidean space, as $V \to 0$ as $R \to \infty$.

\subsubsection{The function $f$ and one-form $\omega$}
The full five-dimensional metric is written in the following form, 
\begin{align}
&ds^2 = f^2 (d\tau + \omega)^2 + f^{-1} ds^2_4 & &\\
&ds^2_4 = V^{-1} (d \Psi  + A)^2 + V ds^2_3 & &\\
&ds^2_3 =dy_1^2 + dy_2^2 + dy_3^2,
\end{align}
with 
$\omega = \omega_\phi d\phi + \omega_\psi d\psi  
$ and $\star_3 dA = dV$.

Now we write the various functions appearing in the metric in terms of harmonic functions
 closely following the notation in the Lorentzian signature \cite{Gauntlett:2002nw, Bena:2005va, Bena:2007kg}. The fact that we are able to do this is the third hint that the solution is supersymmetric. We introduce three more harmonic functions $K, L, M$  
\begin{align}
&K =  k \left( \frac{1}{R_+} - \frac{1}{R_-}\right), & 
&L =1 +   \frac{l_+}{R_+} +  \frac{l_-}{R_-},& 
&M = \frac{m_+}{R_+} +  \frac{m_-}{R_-},&
\end{align}
with coefficients, 
\begin{align}
&k = \frac{q}{4 \sqrt{d^2+q}}, & & \\
&l_+ = \frac{q}{8}+ \frac{dq}{8 \sqrt{d^2+q}}, &
&l_- = \frac{q}{8}-\frac{dq}{8 \sqrt{d^2+q}},  & \\
&m_+ = \frac{qd}{16} + \frac{q^2}{32 \sqrt{d^2+q}}+\frac{d^2 q}{16 \sqrt{d^2+q}}, & 
&m_- = \frac{qd}{16} - \frac{q^2}{32 \sqrt{d^2+q}} - \frac{d^2 q}{16 \sqrt{d^2+q}}.&  
\end{align}
Note that the $K$ function has  zero net charge while the $L$ and $M$ functions have net charges $\frac{q}{4}$ and $\frac{q d}{8}$, respectively. 
A calculation shows that
\be
\omega = \mu ( d \Psi + A) + \omega_3,
\ee
where
\be
\mu = \frac{K^3}{V^2} - \frac{3}{2} \frac{K L}{V} + M,
\ee
and
\be
\star_3 d \omega_3 = V dM - M dV - \frac{3}{2} \left( K dL - L dK \right).
\ee
Clearly these equations are similar to one written in \cite{Gauntlett:2002nw, Bena:2005va, Bena:2007kg}, but some minus signs are different, as the solutions we are working with are solutions of Euclidean supergravity. The function $f$ has the form, 
\be
f^{-1} = L - \frac{K^2}{V}.
\ee

\subsubsection{The vector field}
The vector field supporting the five-dimensional Euclidean $m=q$ Cvetic-Youm solution can also be written in terms of the harmonic functions introduced above as
\be
A_E =  \sqrt{3}\left( d\tau -  f (d\tau + \omega) -   \frac{K}{V}(d \Psi + A) + \xi \right),
\ee
where
$
\star_3 d \xi = d K.
$ 
We have also separately checked that configurations specified by the four (arbitrary) harmonic functions $V, K, L, M$ solves equations to Euclidean minimal supergravity. The order of coordinates consistent with our conventions is $(\tau, \Psi, y_1, y_2, y_3)$.
\subsection{Killing spinors}
The Killing spinor equation \eqref{Killing-spinor-eq-5d}
can be solved straightforwardly for the general class of solutions discussed above. For the choice of the frame fields,
\begin{align}
e^1 &= f (dt + \omega), &
e^2 &= f^{-1/2} V^{-1/2}  (d \Psi + A), &\\
e^3 &=f^{-1/2} V^{1/2}  d y_1,& 
e^4 &= f^{-1/2} V^{1/2}  d y_2,&
e^5 &= f^{-1/2} V^{1/2}  d y_3,&
\end{align}
we have,
\be
\e =
f^{1/2} \left( \begin{array}{c}
 \e_0  \\
 \e_0
\end{array}
\right), \label{eq:spinor_5d}
\ee
where $\e_0$ is a constant two-component complex spinor: $\partial_\mu
\e_0 = 0$. 
For the $m=q$ Euclidean Cvetic-Youm solution $f \to 1$ as $r \to \infty$; i.e., the Killing spinors are normalisable.  Under $2 \pi$ rotation in the $(y_1,y_2)$ plane, $\e$ goes to $\mathrm{exp}\left[ (2 \pi) \cdot 2 \cdot  \frac{1}{4} \Gamma^{34} \right] \e= -\e$. Thus, the Killing spinors satisfy the expected periodicity properties. This implies that  the Killing spinors satisfy  expected periodicity properties is the original coordinates too. To see this, we first note that
\begin{align}
\varphi &=  \phi + \psi, &
\Psi &= \psi - \phi,
\end{align}
Therefore, 
\be
(\phi,  \psi, \tau ) \equiv (\phi + 2 \pi, \psi , \tau) \implies ( \varphi,  \Psi, \tau ) \equiv (\varphi + 2 \pi, \Psi - 2 \pi, \tau),
\label{identifications_5d_2_1}
\ee
and
\be
( \phi,  \psi, \tau ) \equiv (\phi , \psi+ 2 \pi, \tau) \implies ( \varphi,  \Psi, \tau ) \equiv (\varphi + 2 \pi, \Psi+ 2 \pi, \tau).
\label{identifications_5d_2_2}
\ee
Since the Killing spinors do not depend on the $\tau$ and $\Psi$ coordinates, 
as far as the action of  identification \eqref{identifications_5d_2_1} or \eqref{identifications_5d_2_2} on the Killing spinors is concerned, it simply correspond on $2 \pi$ rotation in the $(y_1,y_2)$ plane.

Identification \eqref{identifications_5d} in $( \varphi,  \Psi, \tau ) $ coordinates is,
\bea
( \varphi,  \Psi, \tau ) &\equiv& ( \varphi + \beta \Omega_\psi + \beta \Omega_\phi,  \Psi + \beta \Omega_\psi -\beta \Omega_\phi, \tau + \beta) \\ 
&\equiv& ( \varphi + 2 \pi,  \Psi + \beta \Omega_\psi -\beta \Omega_\phi, \tau + \beta) 
 \label{identifications_5d_2_3}
\eea
where in going from the first line to the second line we have used  the relation $\beta (\Omega_\phi + \Omega_\psi) = 2 \pi$. This relation was noted just after \eqref{identifications_5d}. Once again, since the Killing spinors do not depend on the $\tau$ and $\Psi$ coordinates, identification \eqref{identifications_5d_2_3} simply corresponds on $2 \pi$ rotation in the $(y_1,y_2)$ plane.

\section{Conclusions}
\label{sec:conclusions}

In this paper, we presented the Euclidean Cvetic-Youm solution saturating the BPS bound in the Gibbons-Hawking form. The Gibbons-Hawking form allowed us to exhibit the Killing spinors in  five-dimensional minimal supergravity.  We also expanded on the previous discussions of Killing spinors for the Euclidean Kerr-Newman solution saturating the BPS bound. We saw that for both these cases, the total charge gets divided into two harmonic sources on three-dimensional flat base space. In the four-dimensional case two harmonic functions are sufficient to bring out this physics. In the five-dimensional case, four harmonic functions featured in our analysis.

Our discussion can be generalised in several directions.  The Euclidean solutions we discussed are saddles for the Euclidean gravitational path integral that is related to the supersymmetric index. There are other saddles. For instance, multi-center metrics. A general open problem is how to decide which saddles to include in the gravitational path integral. Supersymmetric black holes embedded in set-ups where precision counting of black hole microstates in known can possibly shed light on this general question, as in these cases we know what answers to expect from the microscopic side. At the very least, one can explore the physics of multi-center metrics (on-shell action, smoothness conditions, etc), extending the earlier discussions \cite{Whitt:1984wk, Yuille:1987vw}, in an attempt to understand their contributions to the gravitational path integral. 

To our knowledge, a classification of supersymmetric solutions of Euclidean five-dimensional minimal supergravity has not been completed. Roughly two decades ago, the classification of Lorentzian supersymmetric solutions was crucial \cite{Gauntlett:2002nw}, which led to the discovery of the supersymmetric black rings \cite{Elvang:2004rt} and of the supersymmetric bubbling solutions \cite{Bena:2005va}. Understanding the spectrum of supersymmetric solutions of Euclidean supergravity will certainly be a step towards understanding their role in the gravitational path integral. 

At some point, we would like to extend our analysis to include vector multiplets in five-dimensions. This will allow us to make progress towards establishing supersymmetry of Euclidean Cvetic-Youm solution saturating the BPS bound with three independent charges. 
 It would also be interesting to understand the relation of our analysis to that of appendix A of \cite{Cabo-Bizet:2018ehj}. 
We hope to report on this in the near future.

\subsection*{Acknowledgements}
We thank James Lucietti and Guillaume Bossard for email correspondence and Ashoke Sen for  encouraging us to pursue this computation. It is a pleasure to also thank Chandramouli Chowdhury and P Shanmugapriya  for interesting and useful discussions. We thank Guillaume Bossard for reading through an earlier version of the draft. We also thank an anonymous referee for insightful comments, which led to improvement in the presentation of the results.
\appendix
\section{Gamma matrix conventions}
\label{app:gamma}

We work with the following representation of Euclidean gamma matrices,
\be
\G^a =
\left(\begin{array}{cc}
0 & -i \s^a \\
i \tilde \s^a & 0
\end{array}\right),
\ee
where $\s^a = (i, \vec \tau \,)$ and $\tilde \s^a = (-i, \vec \tau \,)$. Here
$\vec \tau \,$ are
the Pauli matrices. 
In five-dimensions we also add
\be
\G^5 = \G^1
\G^2 \G^3 \G^4.
\ee
The gamma matrices satisfy $\{\G^a,\G^b\} = 2
\d^{ab}$ for $a, b = 1,\ldots,5$. We define
\be
\G^{ab} \equiv \frac{1}{2} \left[\G^a, \G^b \right] =
\left(\begin{array}{cc}
\s^{ab} & 0 \\
0 & \tilde \s^{ab}
\end{array}\right),
\ee
where $\s^{ab} = \frac{1}{2} \left[\s^a \tilde \s^b - \s^b \tilde
\s^a
  \right]$ and $\tilde \s^{ab} = \frac{1}{2} \left[\tilde \s^a \s^b - \tilde
  \s^b \s^a  \right]$.

Throughout the work, we use Dirac spinors. For the case of Euclidean four-dimensional $\mathcal{N}=2$ supergravity as well as Euclidean five-dimensional minimal supergravity one often also uses symplectic Majorana spinors, which are equivalent to Dirac spinors.  To see this consider symplectic Majorana spinors $\psi^i$ that satisfy,
\be
(\psi^i)^C=\varepsilon_{ij}\psi^j,
\ee
where $i,j=1,2$, and $\varepsilon_{ij}$ is the two dimensional Levi-Civita tensor with $\varepsilon_{12}=1$. In Euclidean signature the Dirac conjugate of a spinor is its hermitian conjugate, and the charge conjugation operation acts as \cite{deWit:2017cle},
\be
\chi^C=C^{-1}\chi^*,
\ee
where $\chi$ can be any kind of spinor. In four Euclidean dimensions the charge conjugation matrix $C$ is antisymmetric and unitary and satisfies
\be
C\,\Gamma^a C^{-1}=-(\Gamma^a)^{\rm T}, \quad a=1,\ldots,4, \qquad\qquad C\,\Gamma^5 C^{-1}=(\Gamma^5)^{\rm T}; \label{eq:A.6}
\ee
in the representation above one can take $C=\Gamma^1\Gamma^3$. (In five dimensions one uses instead a matrix with $C\,\Gamma^A C^{-1}=+(\Gamma^A)^{\rm T}$ for $A=1,\ldots,5$, e.g.\ $C=\Gamma^2\Gamma^4$; the remaining statements of this paragraph are unchanged.) Using these properties, one can show that,
\begin{align}
\left(\chi^C\right)^C=-\chi.
\end{align}
This is true whenever a symplectic Majorana spinor exists in any spacetime dimension or signature. Given a Dirac spinor $\psi$, we can then construct symplectic Majorana spinors via,
\begin{align}
\psi^1&=\frac{1}{2}(\psi-\psi^C),&
\psi^2&=\frac{1}{2}(\psi+\psi^C),
\end{align}
which satisfy $(\psi^1)^C=\psi^2$, $(\psi^2)^C=-\psi^1$ and $\psi^1+\psi^2=\psi$.

\section{Comparison with  ACS notation \cite{Anupam:2023yns}}
\label{app:ACS}

In appendix A of \cite{Anupam:2023yns}, Anupam, Chowdhury, Sen (ACS) conjectured that their metric (A.20) is supersymmetric.\footnote{Together with other fields which they did not write, but whose form can be
obtained from \cite{Cvetic:1996xz}.} As written, the conjecture applies to U(1)$^3$ supergravity. We restrict ourselves to minimal supergravity. The fields relevant for minimal supergravity are  obtained by setting the three independent charges equal. We use little $q$ notation: $Q^{(1)} = Q^{(2)} = Q^{(3)} = Q =q$. Furthermore, we introduce rotation parameters $a$ and $b$ to parameterise their $J_L$ and $J_R$ as follows:
\bea
J_L &= (a-b)q, \qquad J_R &= 3(a+b)q.
\eea
Note the factor of 3 in the definition of $J_R$. Furthermore, $\Delta_{\mathrm{there}} = \Delta_{\mathrm{here}}^3$. $ \Delta_{\mathrm{here}}$ is our $\Delta$. It takes the form,
\be
\Delta = \rho_{ACS}^2 + Q + \frac{J_L J_R  \cos(2 \theta) }{6 Q^2} =\rho_{ACS}^2 + q + \frac{1}{2} (a^2 - b^2) \cos(2 \theta),
\ee
where $\rho_{ACS}$ is the radial coordinate used in writing the metric there. 
Upon shifting of the radial coordinate as
\be
\rho_{ACS}^2 = r^2 + \frac{1}{2} (a^2 + b^2 - 2 q),
\ee
the \underline{Lorentzian} ACS metric takes the Chong-Cvetic-Lu-Pope (CCLP) form 
\eqref{metric_5d_first}--\eqref{metric_5d_last}. For the ease of comparison, we also write the simplified metric from the ACS paper in the ACS notation, 
\bea
g_{tt}\!\!\!&=& -\frac{6 \rho ^2 Q^2 + J_L J_R \cos^2(2 \theta )}{36
   \Delta ^2 Q^4},\\
g_{\rho \rho}\!\!\! &=&\frac{36 \Delta  \rho ^2 Q^4}{36 \rho ^4 Q^4- J_R^2 \left(J_L^2-4
   Q^3\right)} ,\\
g_{\theta \theta}\!\!\! &=& \Delta, \\
g_{t \phi}\!\!\! &=&- \frac{\sin^2 \theta}{12 Q^2 \Delta^2} \left( 6 \rho^2 Q^2
   (J_L+J_R) + J_L J_R (J_L+J_R) \cos (2 \theta ) +4 J_R Q^3 \right), \\
g_{t \psi}\!\!\! &=& -\frac{\cos^2 \theta}{12 Q^2 \Delta^2} \left( 6 \rho^2 Q^2
   (-J_L+J_R) + J_L J_R (-J_L+J_R) \cos (2 \theta ) +4 J_R Q^3 \right),\\
g_{\phi\psi}\!\!\!&=& \frac{\sin ^2 \theta \cos ^2 \theta }{108 \Delta ^2 Q^3}
\left(3 Q^2 \left(9 J_L^2
   Q+J_R^2 \left(4 \rho ^2+3 Q\right)\right)+2 J_L J_R^3
   \cos (2 \theta )\right), \\
g_{\phi\phi}\!\!\! &=&\frac{\sin ^2\theta}{216 \Delta ^2 Q^6} 
\bigg[(J_L J_R^3 \cos ^2(2 \theta )
   \left(J_L^2-2 Q^3\right)+24 J_L^2 J_R^2 \rho ^2 Q^2 \cos
   ^4\theta \nn \\
  && \qquad + ~ 6 \bigg(-Q^5 \sin ^2\theta \left(9 J_L^2 Q-J_R^2
   \left(4 \rho ^2+3 Q\right)\right)-J_L^2 J_R^2 \rho ^2 Q^2 \nn \\
   & & 
\qquad -~6
   J_L J_R \rho ^4 Q^4+36 Q^9+108 \rho ^2 Q^8+108 \rho ^4 Q^7  +36
   \rho ^6 Q^6\bigg) \nn \\ 
   & & \qquad +~2 J_L J_R^2 Q^3 \cos (2 \theta ) (9
   J_L+J_R)+36 J_L J_R Q^4 \cos ^2\theta  \left(4
   \rho ^4+3 Q^2+6 \rho ^2 Q\right)\bigg], \nn  \\ \\
g_{\psi\psi}\!\!\! &=&\frac{\cos ^2\theta}{216 \Delta ^2 Q^6} 
\bigg[
-J_L^3 J_R^3 \cos ^2(2 \theta )-18 J_L^2 J_R^2 Q^3
   \cos (2 \theta )
   \nn \\ && 
\qquad   + ~6 Q^2 \bigg(4 J_L^2 J_R^2 \rho ^2 \sin
   ^4 \theta-J_L^2 J_R^2 \rho ^2-6 J_L J_R Q^2 \sin
   ^2\theta  \left(4 \rho ^4+3 Q^2+6 \rho ^2 Q\right)
   \nn \\ &&   
 \qquad  + ~6 J_L J_R
   \rho ^4 Q^2+36 Q^7+108 \rho ^2 Q^6+108 \rho ^4 Q^5+36 \rho ^6 Q^4\bigg)
   \nn \\ && 
\qquad   + ~ 2 Q^3 \cos ^2\theta \left(3 Q^2 \left(J_R^2 \left(4 \rho ^2+3
   Q\right)-9 J_L^2 Q\right)+2 J_L J_R^3 \cos (2 \theta
   )\right)\bigg],
\eea
where $\rho = \rho_{ACS}$. 

\section{Euclidean $\mathcal{N}=2$ supergravity from superconformal tensor calculus}
\label{app:conventions}
In this appendix, we  derive  pure $N=2$ Euclidean supergravity in four dimensions using the superconformal formalism introduced in \cite{deWit:2017cle}. Of interest to us is the bosonic part of the supergravity action that gives the equations of motion and the Killing spinor equation. 

Similar to Lorentzian $\mathcal{N}=2$ supergravity, we can construct Euclidean supergravity from the superconformal approach by using a vector multiplet as the first compensator and the second compensator can  either be a tensor multiplet, or a nonlinear multiplet or a hypermultiplet~\cite{deWit:1980lyi,deWit:1982na}. We will work with a hypermultiplet compensator. 

The bosonic fields of the Euclidean four dimensional $N=2$ vector multiplet are the vector field $W_\mu$, two real scalars $X_+,X_-$ and a triplet of pseudo-real fields $Y_{ij}$. As we are interested in pure supergravity, we will consider the case of a single vector multiplet. Prepotential in this case is given by,
\begin{align}
	F_\pm=\frac{X_\pm^2}{2}.
\end{align}
Therefore, $\frac{\partial F_\pm}{\partial X_\pm}=X_\pm$, $\frac{\partial^2 F_+}{\partial X_+^2}=1$, $\frac{\partial^2 F_-}{\partial X_-^2}=1$. To obtain Euclidean supergravity, we can break the SO$(1,1)$ $R-$symmetry and dilatation symmetry by using the gauge fixing conditions,
\begin{align}
	X_+=X_-=\frac{1}{2},
\end{align}
where we have also implicitly set the units. Conformal boosts are gauge fixed by setting $b_\mu = 0$. On implementing these gauge conditions, the bosonic action for the vector multiplet obtained from the chiral density formula is given as,
\begin{align}
	e^{-1}\mathcal{L}_V=-\frac{1}{2}A_\mu A^\mu-\frac{1}{2}\left(-\frac{R}{6}-D\right)-\frac{1}{8}F\cdot T+\frac{1}{128}T\cdot T+\frac{1}{4}F\cdot F+  \frac{1}{4}Y_{ij}Y^{ij},
\end{align}
where $A_\mu$ is the $R-$symmetry gauge field and $T_{ab}, D$ are auxiliary fields from the Weyl multiplet. Equations of motion for $A_\mu$, $Y_{ij}$ and $T_{ab}$ lead to,
\begin{align}
	A_\mu&=0,&
	Y_{ij}&=0,&
	T_{ab}&=8F_{ab}.
\end{align}
The action then becomes,
\begin{align}
	e^{-1}\mathcal{L}_V=\frac{1}{2}\left(\frac{R}{6}+D\right)-\frac{1}{4}F\cdot F.
\end{align}

We need to deal  with the problematic term proportional to the Weyl multiplet field $D$, and we  also need to gauge fix the chiral SU(2) symmetry. For this, we use the hypermultiplet as the second compensator. 

The bosonic fields of the Euclidean $N=2$ hypermultiplet are four real scalars $A_i^\alpha$ where $\alpha=1,2$ and the $R-$symmetry index $i=1,2$. We will gauge fix the chiral SU(2) symmetry by using,
\begin{align}
	A_i^\alpha=\delta_i^\alpha e^u,
\end{align}
similar to the Lorentzian case \cite{deWit:1980lyi}. By using all the gauge fixing conditions above, upto total derivatives, the bosonic part of the hypermultiplet action reads,
\begin{align}
	e^{-1}\mathcal{L}_H=-\frac{1}{8}e^{2u}V^{ak}{}_jV_a{}^j{}_k-e^{2u}\left(-\frac{R}{6}+\frac{D}{2}\right),
\end{align}
where $V_a{}^i{}_j$ is the chiral SU(2) symmetry gauge field. Its equation of motion gives,
\begin{align}
	V_a{}^i{}_j=0.
\end{align}
We can get the full action for the Euclidean supergravity by taking a linear combination of the two actions,
\begin{align}
	e^{-1}\mathcal{L}=\frac{R}{12}(1+2e^{2u})-\frac{1}{4}F\cdot F+\frac{D}{2}(1-e^{2u}).
\end{align}
Equation of motion for the auxiliary field $D$ now gives $u=0$. Therefore the action is,
\begin{align}
	e^{-1}\mathcal{L}=\frac{R}{4}-\frac{1}{4}F\cdot F.
\end{align}

To obtain the Killing spinor equation, note that the variation of gravitini in conformal supergravity is given by,
\begin{align}
	\delta \psi_\mu^i&=2\mathcal{D}_\mu\epsilon^i+\frac{i}{16}\Gamma\cdot T \Gamma_\mu \epsilon^i-i\Gamma_\mu \eta^i,
\end{align}
where using the gauge fixing we have done above,
\begin{align}
	\mathcal{D}_\mu\epsilon^i=\partial_\mu\epsilon^i+\frac{1}{4}\Gamma\cdot\omega(e)_\mu\epsilon^i,
\end{align}
where $\omega(e)_\mu{}^{ab}$ is the spin-connection determined in terms of the vielbein via the first structure equation.  It is possible that the $Q$-transformation of $\psi_\mu^i$ needs to be accompanied by a compensating $S$-transformation when we gauge fix $S$-supersymmetry.
This will turn out to be not the case since we use fermions of the vector multiplet $\Omega_\pm^i$ to gauge fix the $S$-supersymmetry, and on support of the gauge conditions and auxiliary field solutions derived above, the $Q$-variation of $\Omega_\pm^i$ vanishes. Therefore the supersymmetric variation of the gravitini in Euclidean supergravity reads,
\begin{align}
	\delta \psi_\mu^i = 2\left[\partial_\mu\epsilon^i+\frac{1}{4}\Gamma\cdot\omega(e)_\mu\epsilon^i+\frac{i}{4}\Gamma\cdot F\Gamma_\mu\epsilon^i\right].
\end{align}
In the above, the spinor $\epsilon^i$ is a symplectic Majorana spinor \cite{deWit:2017cle}, $(\epsilon^i)^C=\varepsilon_{ij}\epsilon^j$, with the charge conjugation defined in appendix \ref{app:gamma}. On the solution $A_\mu=0$ and $V_\mu{}^i{}_j=0$, so the condition $\delta\psi_\mu^i=0$ does not mix $\epsilon^1$ and $\epsilon^2$ and takes the same form for both.  We can therefore write the Killing spinor equation in terms of a single Dirac spinor as,
\begin{align}
\partial_\mu\e+\frac{1}{4}\Gamma\cdot\omega(e)_\mu\e+\frac{i}{4}\Gamma\cdot F\Gamma_\mu\e=0,
\end{align}
where $\e=\epsilon^1+\epsilon^2$.  Moreover, since $\omega(e)_\mu{}^{ab}$ and $F_{ab}$ are real and $C\Gamma^aC^{-1}=-(\Gamma^a)^{\rm T}$, cf.~\eqref{eq:A.6}, the charge conjugate $\e^C=C^{-1}\e^*$ of any solution $\e$ of this equation is again a solution, with the same sign of the $\frac{i}{4}\Gamma\cdot F\Gamma_\mu$ term.
Conversely, given a solution $\e$, the pair $\epsilon^1=\frac{1}{2}(\e-\e^C)$, $\epsilon^2=\frac{1}{2}(\e+\e^C)$ is a symplectic Majorana solution. This Killing spinor equation matches that of \cite{Dunajski:2006vs,Dunajski:2010uv}.

\end{document}